\documentclass[sigplan,screen,9pt]{acmart}
\renewcommand\footnotetextcopyrightpermission[1]{} 
\usepackage{balance}
\usepackage{subcaption}
\usepackage{caption}
\usepackage[normalem]{ulem}
\usepackage[table,xcdraw]{xcolor}
\usepackage{tcolorbox}
\usepackage{balance}
\usepackage{hyperref}
\usepackage{url}
\usepackage{fancyhdr}
\usepackage{dirtytalk}
\usepackage{graphicx}
\usepackage{multirow}
\usepackage{multicol}
\usepackage{subcaption}
\usepackage{algorithm}
\usepackage{algpseudocode}
\usepackage{amsmath}
\usepackage{wrapfig}
\newcommand{\blue}[1]{\textcolor{black}{#1}}

\newcommand{\abstractword}[1]{\gdef\@abstractword{#1}}
\begin{document}

\title{
COBRA: \underline{C}atastr\underline{o}phic \underline{B}it-flip \underline{R}eliability \underline{A}nalysis of State-Space Models
}
\pagestyle{plain}
\author{\large Sanjay Das$^{1*}$, Swastik Bhattacharya$^1$, Shamik Kundu$^2$, Arnab Raha$^2$, Souvik Kundu$^2$,  and Kanad Basu$^3$\\
\textit{$^1$University of Texas at Dallas, USA}; \textit{$^2$Intel Corporation, USA}; \textit{$^3$Rensselaer Polytechnic Institute, USA} \\
$^*$Corresponding author: Sanjay Das (Email: \textit{sanjay.das@utdallas.edu})}
\begin{abstract}


State-space models (SSMs), exemplified by the Mamba architecture, have recently emerged as state-of-the-art sequence-modeling frameworks, offering linear-time scalability together with strong performance in long-context settings. Owing to their unique combination of efficiency, scalability, and expressive capacity, SSMs have become compelling alternatives to transformer-based models, which suffer from the quadratic computational and memory costs of attention mechanisms. As SSMs are increasingly deployed in real-world applications, it is critical to assess their susceptibility to both software- and hardware-level threats to ensure secure and reliable operation.

Among such threats, hardware-induced bit-flip attacks (BFAs) pose a particularly severe risk by corrupting model parameters through memory faults, thereby undermining model accuracy and functional integrity. To investigate this vulnerability, we introduce COBRA, the first BFA framework specifically designed to target Mamba-based architectures. Through experiments on the Mamba-1.4b model with LAMBADA benchmark—a cloze-style word- prediction task—we demonstrate that flipping merely a \textbf{single} critical bit can catastrophically reduce accuracy from 74.64\% to 0\% and increase perplexity from 18.94 to 3.75 × 10$^6$. These results demonstrate the pronounced fragility of SSMs to adversarial perturbations. 
The framework is open-sourced at \url{https://anonymous.4open.science/r/RAMBO-DA22}.
  
\end{abstract}
\setcopyright{none}
\settopmatter{printacmref=false} 




\maketitle

\section{Introduction}
The increasing popularity of natural language processing (NLP) models has fundamentally expanded the capabilities of Artificial Intelligence (AI), demonstrating remarkable proficiency in generating human-like text, interpreting nuanced context, and executing complex reasoning tasks~\citep{xu2024survey}. These advancements have not only reshaped natural language processing but have also extended AI applications into diverse fields such as computer vision and scientific research, heralding a new era of AI-driven solutions~\citep{chang2024survey, xu2024survey}.  State-space models (SSMs), such as Mamba, have emerged as leading sequence-modeling architectures, offering linear-time scalability and strong performance on long-context tasks~\cite{Mamba, Mamba2}. 
Therefore, they have garnered attention as highly attractive alternatives to conventional transformer-based large language models due to their unique combination of efficiency, scalability, and representational strength. Unlike transformers, whose attention mechanism incurs quadratic computational and memory costs, SSMs operate in linear time, enabling fast processing of extremely long sequences~\cite{Mamba}.
As SSMs continue to be integrated into real-world systems at an accelerated pace, it becomes increasingly important  to analyze their vulnerability against both software-based and hardware-based threats to ensure their secure and reliable deployment ~\citep{das2024security, das2024attentionbreaker}.

A major concern in the reliability of deep learning models arises from hardware-level attacks such as \emph{bit-flip attacks} (BFAs), which exploit vulnerabilities in memory to corrupt the model’s weight parameters. Such corruption can severely degrade model performance and violate its integrity. For example, BFA methodologies including \emph{DeepHammer} inject faults into DRAM, flipping specific bits in stored weights to impair functionality~\citep{yao2020deephammer}.  
Even with advances in memory technology, recent techniques allow for remote, non-physical bit-flip manipulations, thereby expanding the threat surface available to BFAs~\citep{hayashi2011non, shuvo2023comprehensive}. 
Bit-flip attacks (BFAs) have been studied extensively in the context of conventional deep neural networks (DNN)~\citep{qian2023survey, rakin2019bit, chen2021proflip}. Howver, traditional BFA strategies often require iterative gradient recomputation after each individual bit‐flip. While this is feasible for comparatively small models, it becomes computationally intractable as model size increases ~\citep{rakin2019bit, kundubit}. Recent work has begun to expose severe vulnerabilities of transformer-based LLMs to BFAs via alternative strategies, demonstrating that as few as a single bit-flip can catastrophically degrade LLM performance~\citep{das2024attentionbreaker, nazari2024forget}. However, BFA implications for alternative sequence modeling paradigms — in particular structured state-space models (SSMs) — remain largely unexplored.  

State-space architectures such as Mamba implement fundamentally different information-propagation and parameterization mechanisms, trade off recurrence and selectivity for attention mechanisms
~\citep{gu2023Mamba}.  Due to these architectural distinctions, it is not appropriate to assume that attack strategies and defenses developed for transformers transfer directly to SSMs.  
Consequently, the absence of a systematic study of BFAs on SSMs constitutes a substantive gap in the current AI hardware robustness and security literature.
To address this gap, we propose 
the framework ``\emph{COBRA}'' — a first of its kind SSM-aware BFA pipeline. 
The primary contribution of the paper are as follows: 

\begin{itemize}
  \item \blue{We identify and formalize a previously unexplored vulnerability of state-space models (\textit{e.g.}\ Mamba) to bit-flip attacks, and propose COBRA, a first of its kind SSM-aware attack approach, bridging the gap between BFA research and structured sequence models.}
  \item \blue{COBRA leverages the structural properties of Mamba-style SSM layers to prioritize critical parameter regions, adapts gradient-estimation and search heuristics, and uses the resulting perturbation effects to identify minimal bit-flip sets that maximally disrupt model behavior. }
  \item COBRA uncovers a significant vulnerability of Mamba models. A mere \textbf{one bit-flip}
(7.14 × 10$^{-10}$\% of all bits) in Mamba-1.4b, can reduce the  LAMBADA word prediction accuracy from
74.64\% to 0\%, while increasing Wikitext perplexity from 18.94 to 3.75 × 10$^6$. 
\end{itemize}
The rest of the paper is organized as follows: Section \ref{sec:background} provides relevant background information. The threat model is discussed in Section \ref{sec:threat} and the proposed methodology is detailed in Section \ref{sec:methodology}. Section \ref{sec:results} outlines the experimental setup and discusses the results. The concluding remarks are offered in Section \ref{sec:conclusion}.

\section{Background}\label{sec:background}
State-space models such as Mamba are architecturally composed of stacked Mamba blocks, each integrating selective state-space layers and projection components that collectively enable the model’s long-range sequence modeling capabilities.

\subsection{State-Space Dynamics}
A Mamba block operates as a parameter-efficient state-space model (SSM) augmented with convolution, projection, and normalization layers~\cite{Mamba}.  
At each time-step $t$, the latent state $h_t \in \mathbb{R}^{n}$ evolves according to the recurrence
\begin{equation}
    h_{t+1} = (I + \Delta_t A)\, h_t \;+\; \Delta_t B_t x_t,
    \label{eq:ssm-update}
\end{equation}
and produces an output
\begin{equation}
    y_t = C_t h_t + D x_t,
    \label{eq:ssm-output}
\end{equation}
where:
\begin{itemize}
    \item $A \in \mathbb{R}^{n \times n}$ is the state-transition matrix, parameterized for stability and fixed after training,
    \item $B_t, C_t \in \mathbb{R}^{n}$ are input-dependent write and read vectors,
    \item $D \in \mathbb{R}$ is a fixed skip coefficient,
    \item $\Delta_t \in \mathbb{R}_{+}$ is a token- and channel-dependent step size controlling the effective timescale,
    \item $x_t \in \mathbb{R}$ is the projected token input.
\end{itemize}

This recurrence is analogous to a recurrent neural network (RNN), but with structured dynamics that can be parallelized efficiently via diagonalization and convolutional scan operations.

\subsection{Parameterization via Projection Layers}
Let $m$ denote the model embedding dimension, $n$ the latent state dimension per channel, and $r$ the low-rank dimension used for step-size generation.  
The projection pipeline operates as follows:
\begin{align}
    u_t &= W_{\mathrm{in}} x_t \in \mathbb{R}^{c}, \qquad c \approx m, \\
    p_t &= W_{\mathrm{proj}} u_t \in \mathbb{R}^{2n + r}, \\
    p_t &= \big( B^{\mathrm{raw}}_t, C^{\mathrm{raw}}_t, \Delta^{\mathrm{low}}_t \big), \\
    \Delta_t &= \operatorname{softplus}(W_{\Delta} \Delta^{\mathrm{low}}_t) \in \mathbb{R}^{c}.
\end{align}
Here:
\begin{itemize}
    \item $W_{\mathrm{in}} \in \mathbb{R}^{m \times c}$ projects model embeddings into an intermediate space,
    \item $W_{\mathrm{proj}} \in \mathbb{R}^{c \times (2n+r)}$ generates raw seeds for $B_t$, $C_t$, and the low-rank step-size representation,
    \item $W_{\Delta} \in \mathbb{R}^{r \times c}$ expands the low-rank $\Delta^{\mathrm{low}}_t$ into a per-channel step-size vector.
\end{itemize}

Thus, although $A$ and $D$ are fixed parameters of the model, the effective dynamics are governed by \emph{input-dependent} $B_t$, $C_t$, and $\Delta_t$ that vary across tokens. Therefore, these unique structural characteristics and projection parameterization of Mamba models must be considered in the development of efficient bit-flip attack strategies.


\section{Threat Model}\label{sec:threat} 
\begin{figure}[t]
\centerline{\includegraphics[width=1.05\columnwidth]{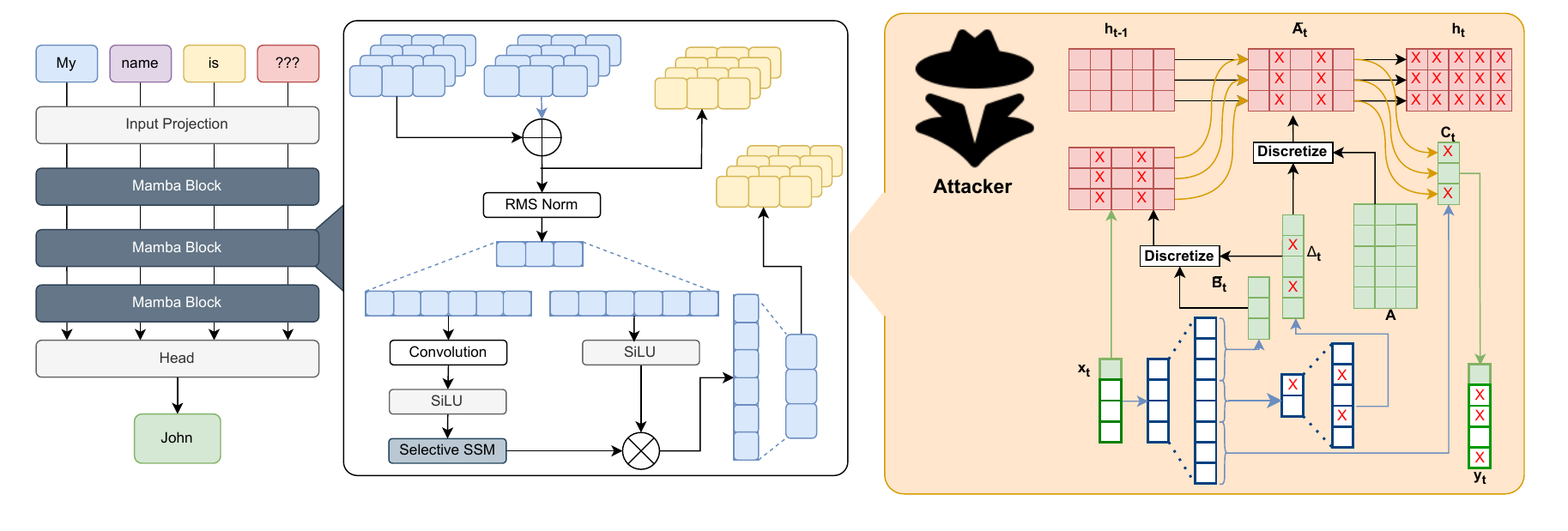}}
\caption{Bit-flip attack on state space models like Mamba.}
\label{fig:ssm}
\vspace{-3mm}
\end{figure}

The proliferation of large-scale NLP models has heightened concerns over security risk, including backdoor and inference-time attacks~\citep{saha2020hidden, xie2019dba, goodfellow2014explaining, moosavi2017universal}. Beyond these vectors, a more insidious threat arises from direct manipulation of model parameters~\citep{breier2018practical}. Under this threat model, an adversary with low-level memory access can alter stored weights of deployed models to induce malicious behavior. Hardware-based attacks, such as RowHammer~\citep{kim2014flipping} and Laser Fault Injection~\citep{selmke2015precise}, enable such bit-level perturbations to critical model parameters. These bit-flips inject critical errors into the model’s computational flow, propagating and compounding across layers and blocks, ultimately producing erroneous outputs, as illustrated for SSMs such as Mamba in Figure~\ref{fig:ssm}.
The risk is further amplified in Machine Learning as a Service (MLaaS) settings, where shared hardware resources can expose models to co-residency and cross-process vulnerabilities~\citep{das2024attentionbreaker}.

We categorize the threats into four levels based on the extent of knowledge an attacker can extract from the victim. A \textbf{full-knowledge attack} occurs when the attacker has complete information about the target model, training data and its defenses, enabling them to devise highly optimized attack strategies \cite{yarom2014flush+, kayaalp2016high}. In a \textbf{white-box attack}, attackers have full knowledge of the model parameters but lack direct memory access, training data, or its defenses, enabling them to execute fault-injection attacks such as Rowhammer \cite{madry2017towards, rakin2019bit}. A \textbf{gray-box attack} involves partial knowledge of the model, allowing adversaries to exploit known vulnerabilities based on available information \cite{xiang2020side,zhang2018generalized}. Finally, in a \textbf{black-box attack}, attackers have no direct access to the target model’s architecture or parameters. Attackers can only observe model outputs to infer model properties or extract sensitive information \cite{papernot2017practical,mahmood2021back}. 


In this work, we consider both white-box and gray-box threat models. The white-box setting applies because the targeted models are open-sourced, granting full access to their architecture and parameters. The gray-box setting reflects scenarios in which the adversary has only partial visibility into model internals. Within this setting, we focus on untargeted attacks, which aim to induce graceless degradation in overall model performance rather than targeting specific outputs. Such attacks are particularly insidious, as they compromise accuracy across a broad range of inputs while avoiding distinct failure patterns, making them more challenging to detect and defend against than targeted attacks \citep{codematch, qian2023survey, nazari2024forget}. 
Therefore, COBRA strategically employs an untargeted attack to maximize disruption and degrade model performance. 
\emph{This attack also fundamentally differs from denial of service attacks, which aim to overwhelm system resources to render services unavailable.} 

\section{Proposed COBRA Methodology}\label{sec:methodology}
In this section, we present a theoretical sensitivity analysis followed by a detailed description of the proposed attack framework.

\subsection{Theoretical Vulnerability Analysis}\label{sec:theory}
Here, we present parameter-ranking procedure to identify plausible targets for the COBRA attack. We rank the SSM parameters' based on their importance 
 as follows:

\begin{itemize}
    \item \textbf{Transition backbone (A)}: This component determines how information evolves over time within each channel. Because (A) is applied at every timestep and across all layers, its parameters strongly influence the model’s stability and its ability to capture long-range dependencies.

    \item \textbf{Projection seeds}: These seeds generate the raw projection vectors ($B_{\mathrm{raw}}$), ($C_{\mathrm{raw}}$), and the low-rank update ($\Delta_{\mathrm{lowrank}}$). If the seeds carry insufficient information, the model cannot effectively adapt to the input context.
    \item \textbf{Seed expansion parameters}: These parameters convert the low-rank ($\Delta$) seed into a detailed per-channel timescale vector. Poor calibration can produce unusable timescales—values that are too small prevent meaningful updates, while values that are too large destabilize the model.
    \item \textbf{Read/write vectors ($B_t$) and ($C_t$)}: Derived from the seeds, these vectors govern how new information is written into the model’s state and how existing information is retrieved.

\end{itemize}

From the ranking above, the $A$ modules and the projection seeds emerge as the most critical components. However, projection seeds are input dependent and incur substantial runtime overhead to probe and identify, making them impractical targets for real-time attacks. Consequently, it is more practical to target fixed model parameters such as the $A$ and $D$ modules and the projection layers, which are static and can be reliably identified and exploited.


\subsection{Attack Framework}
In this section, we delineate our attack methodology for impacting Mamba model performance.
We employ the standard cross-entropy loss~\citep{mao2023cross}, computed between the output logits of the model and the corresponding ground-truth token IDs, as a measure of model performance. A lower cross-entropy loss indicates that the model assigns higher likelihood to the correct tokens and thus performs better. Our objective is to identify the most critical subset of parameter bits such that, when flipped, they cause a substantial increase in the cross-entropy loss. This increase directly translates into severe degradation of model performance, thereby highlighting the vulnerability of the model to targeted bit-flip manipulations.
\subsubsection{Proxy for Parameter Sensitivity}\label{sec:proxy}
For performing BFA, it is essential to analyze model parameter sensitivity profiles independently of robustness assumptions. In particular, parameters with larger gradients or higher magnitudes may exhibit amplified sensitivity, whereby perturbations yield disproportionately large effects on the output. Therefore,  a hybrid sensitivity metric is
used upon experiments after being inspired by \cite{das2024attentionbreaker}, that considers both magnitude and gradient
influences to capture the sensitivity profile holistically, and is therefore expressed as:
\begin{equation}
    \mathbf{S} = \alpha \cdot |\nabla \mathbf{W}| + (1 - \alpha) \cdot |\mathbf{W}|
    \label{eq:sensitivity}
\end{equation}
Where $\nabla \mathbf{W}$ refers of parameter gradients, $\mathbf{W}$ refers to the magnitudes and \( \alpha \) is a tunable parameter balancing the importance of magnitude and gradient. 

\subsubsection{Layer Sensitivity Analysis}\label{sec:mot_sensitivity}

Determining critical parameters in Mamba Models is complex due to the size of their parameter space. However, the identification of a sensitive layer is more manageable due to the reduced number of layers compared to the total number of parameters. To quantify layer sensitivity, we sample the top-(k) candidate bit-flips from each layer at a rate ($r$), guided by the hybrid sensitivity score ($\mathbf{S}$). These selected bit-flips are applied, and the resulting model loss ($\mathcal{L}$) is measured to assess the layer’s sensitivity. $k$ is computed as:

\begin{equation}\label{eq:topk}
    k = \texttt{cardinality}(\mathbf{W}^{(l)}) \times \frac{r}{100}
\end{equation}

Here, $\mathbf{W}^{(l)}$ signifies the parameters within layer $l$.

To rank layers, we first quantify each layer’s parameter sensitivity using a scoring function (Eq. \ref{eq:sensitivity}). The resulting scores are ranked in descending order to identify the top-(k) most critical weights (Eq. \ref{eq:topk}). Bit-flip perturbations are then applied to the most significant bits of these weights to maximize deviation. The resulting loss, ($\mathcal{L}^{(l)}$), indicates each layer’s sensitivity. 


\begin{algorithm}[t!]
\caption{Layer Ranking \& Wright Subset Selection}
\label{alg:layer_ranking}
\begin{algorithmic}[1]
\Require Model parameters $\mathbf{W}$, gradients $\nabla\mathbf{W}$, trade-off $\alpha$, 
sub-sampling rate $r$, and number of top layers $n$
\Ensure Sensitivity scores $\mathcal{L}_{sens}$, Critical weight indices $\mathcal{I}_{init}$

\State Initialize sensitivity list $\mathcal{L}_{sens} \gets [\;]$

\ForAll{layers $l \in L$}
    \State $k \gets \lfloor r \times |\mathbf{W}^{(l)}| / 100 \rfloor$
    \State $\mathbf{S}^{(l)} \gets \alpha |\nabla\mathbf{W}^{(l)}| + (1 - \alpha)|\mathbf{W}^{(l)}|$
    \State $I^{(l)}_{hyb} \gets \texttt{TopKIndex}(\mathbf{S}^{(l)}, k)$
    \State $\mathcal{L}^{(l)} \gets \texttt{BFlipLoss}(\mathbf{W}^{(l)}, pos, I^{(l)}_{hyb})$
    \State Append $[\mathcal{L}^{(l)}, l]$ to $\mathcal{L}_{sens}$
\EndFor

\State $\mathcal{L}_{sens} \gets \texttt{SORT}(\mathcal{L}_{sens})$
\State $\mathcal{L}_{top} \gets \texttt{TopN}(\mathcal{L}_{sens}, n)$
\State $\mathcal{I}_{init} \gets [l, I^{(l)}_{hyb}]$ extracted from $\mathcal{L}_{top}$
\end{algorithmic}
\end{algorithm}

Algorithm \ref{alg:layer_ranking} outlines a systematic procedure for layer sensitivity analysis and ranking. The algorithm begins by initializing an empty set, $\mathcal{L}_{sens}$ to store layer sensitivity scores (\textit{line 1}). It uses a function $\texttt{BFlipLoss}$ to calculate the model loss $\mathcal{L}$ when weight perturbations are applied to a specified layer (\textit{line 6}). The function $\texttt{BFlipLoss}$ accepts the parameters $\mathbf{W}^{(l)}$, bit position $pos$, and perturbation indices $I$ (\textit{line 2}) as inputs. 
The computed loss $\mathcal{L}$ is returned. The process iterates over each model layer to evaluate its sensitivity to parameter faults (\textit{lines 2–9}). For each layer $l$, a \textbf{hybrid sensitivity score} $\mathbf{S}^{(l)}$ is computed using the weighted combination of parameter magnitudes and gradient magnitudes (\textit{Eq. \ref{eq:sensitivity}, line 4}). The \texttt{TopKIndex} function then selects the top-$k$ most sensitive weights, forming the index set $I^{(l)}_{\text{hyb}}$ (\textit{line 5}). These indices, together with the layer weights $\mathbf{W}^{(l)}$ and bit position `pos', are passed to the \texttt{BFlipLoss} function, which injects controlled bit-flips, recomputes corresponding model loss. 
The losses are recorded in the sensitivity list $\mathcal{L}_{\text{sens}}$ (\textit{line 7}). After processing all layers, the sensitivity list is sorted, and the top-$n$ most vulnerable layers are identified (\textit{lines 9–10}). Finally, the corresponding weight indices from these layers are extracted to form the \textbf{critical weight indices} $\mathcal{I}_{init}$, which, along with $\mathcal{L}_{\text{sens}}$, constitutes the algorithm’s output (\textit{line 11}).

\subsubsection{Critical Parameter Set Optimization}\label{sec:algo_opt}
The initial set of critical parameters can be large and computationally prohibitive for an efficient BFA, Therefore, it becomes necessary to identify a minimal subset that still preserves the original attack effectiveness. Let the initial set of indices be denoted as ($\mathcal{I}_{\text{init}}$), associated with a baseline attack loss $(L_{\text{orig}} = L(\mathcal{I}_{\text{init}}))$. The optimization objective is to find the smallest subset $(\mathcal{I} \subseteq \mathcal{I}_{\text{init}})$ that maintains the attack performance within a small tolerance:

\begin{equation}
    \min_{\mathcal{I} \subseteq \mathcal{I}_{\text{init}}} |\mathcal{I}| \quad \text{s.t.} \quad L(\mathcal{I}) \ge L_{\text{orig}} - \varepsilon.
\end{equation}

This problem can be reformulated as a combinatorial optimization task over binary selection variables ($z_i \in {0, 1}$), where ($z_i = 1$) indicates inclusion of the (i)-th parameter in the subset. The corresponding formulation becomes:

\begin{equation}
\min_{z_i \in {0,1}} \sum_i z_i \quad \text{s.t.} \quad L(z) \ge L_{\text{orig}} - \varepsilon.
\end{equation}

Since this problem is NP-hard, a continuous relaxation can be applied by allowing ($z_i \in [0, 1]$) and using a differentiable surrogate loss ($\hat{L}(z)$). The relaxed formulation is expressed as:

\begin{equation}
\min_{z \in [0,1]^n} \sum_i z_i + \lambda \cdot \max(0, L_{\text{orig}} - \hat{L}(z) - \varepsilon),
\end{equation}

where $\lambda$ is a regularization parameter that balances sparsity and loss preservation. Although this relaxation enables gradient-based optimization, the underlying loss landscape remains highly nonconvex and computationally expensive to evaluate in large-scale neural models.

To address this issue, a randomized exclusionary heuristic is adopted, as described in Algorithm 2. The algorithm iteratively refines the subset by randomly excluding groups of indices and re-evaluating the loss. In each iteration, a candidate exclusion set ($\Delta$) is selected, where ($|\Delta|$) varies between 1 and ($\lfloor |\mathcal{I}| / 2 \rfloor$). The modified subset is tested as:
$\mathcal{I}' = \mathcal{I} \setminus \Delta,$
and the new loss ($L(\mathcal{I}')$) is compared against the baseline. If the loss satisfies
$L(\mathcal{I}') \ge L_{\text{orig}} - \varepsilon$, the exclusion is accepted, permanently removing those indices from the set. The process continues until no further exclusion satisfies the constraint or the maximum number of iterations is reached.

This exclusionary optimization approach provides a \textbf{computationally efficient}  method for subset reduction. Although it does not guarantee global optimality, it achieves significant reduction in the number of critical indices while maintaining nearly identical attack loss — offering a practical balance between optimization cost and attack efficacy.

\begin{algorithm}[t!]
\caption{Exclusionary Weight Subset Optimization}
\label{alg:exclusionary_optimization}
\begin{algorithmic}[1]
\Require Model parameters $\mathcal{W}_{orig}$, 
 weight indices $\mathcal{I}_{init}$, loss tolerance $\epsilon$, maximum iterations $N_{max}$
\Ensure Reduced index set $\mathcal{I}_{red}$ 

\State $\mathcal{I}_{red} \gets \mathcal{I}_{init}$, $\mathcal{P} \gets \{\}$ \Comment{Track progress per layer}
\State $\mathcal{L}_{orig} \gets \texttt{BFlipLoss}(\mathcal{W}_{orig}, pos, \mathcal{I}_{init})$
\State $\texttt{improved} \gets \text{True}$, $t \gets 0$
    \While{$\texttt{improved}$ and $t < N_{max}$ 
    }
        \State $\texttt{improved} \gets \text{False}$, $t \gets t + 1$
        \For{$i = 1 \ \text{to}\ 100$} \Comment{Try random exclusion patterns}
            \State Randomly exclude $n_{exc} \in [1, |\mathcal{I}
            |/2]$ indices
            \State Form $\mathcal{I}_{test}= \mathcal{I} \setminus \mathcal{I}_{exc}$
            \State $\mathcal{L}_{test} \gets \texttt{BFlipLoss}(\mathcal{W}_{orig}, pos, \mathcal{I}_{test})$
            \If{$\mathcal{L}_{test} \ge \mathcal{L}_{orig} - \epsilon$}
                \State $\mathcal{I} \gets \mathcal{I}_{test}$, $\texttt{improved} \gets \text{True}$
                \State \textbf{break}
            \EndIf
        \EndFor
        \State Record progress $(t, |\mathcal{I}|, \mathcal{L}_{test})$
    \EndWhile
    \State $\mathcal{I}_{red} \gets \mathcal{I}$, $\mathcal{P} \gets$ recorded progress
\State \Return $\{\mathcal{I}_{red}, \mathcal{P}\}$
\end{algorithmic}
\end{algorithm}

Algorithm \ref{alg:exclusionary_optimization} presents the optimization procedure. 
The algorithm begins by initializing the reduced index set ($I_{\text{red}}$) with the initial subset ($I_{\text{init}}$) and recording structures for tracking progress (P) (\textit{lines 1–2}). The baseline loss ($L_{\text{orig}}$) is computed using the \texttt{BFlipLoss} function (\textit{line 2}). At each iteration, random exclusion patterns are explored to identify indices that can be safely removed without violating the predefined loss tolerance ($\epsilon$) (lines 4–6). Specifically, a random subset of indices ($I_{\text{exc}} \subseteq I$) of size up to half of the current subset is excluded to form a test subset ($I_{\text{test}} = I \setminus I_{\text{exc}}$) (\textit{lines 7–8}). The resulting model loss ($L_{\text{test}}$) is then evaluated (line 9).
If the condition ($L_{\text{test}} \geq L_{\text{orig}} - \epsilon $) holds, indicating negligible degradation, the exclusion is accepted and ($I$) is updated accordingly (\textit{lines 10–12}). The process continues until no further improvement is observed or the maximum number of iterations ($N_{\text{max}}$) is reached (\textit{lines 4–16}). Finally, the optimized reduced index set ($I_{\text{red}}$) and the recorded progress (P) are obtained (\textit{lines 17–18}).


\section{Evaluation Results}\label{sec:results}
\subsection{Experimental Setup}
We evaluated COBRA on a diverse set of models, including Mamba and Mamba2, ranging from 370 million and 2.8 billion parameters~\cite{Mamba, Mamba2}. We further assess COBRA on quantized models such as Quamba-1.4b-w8a8 (8-bit weights and 8-bit activations), Quamba-1.4b-w4a16, Quamba-2.8B-w8a8, and Quamba-2.8b-w4a16 to evaluate its effectiveness under low-precision settings.
Model performance was assessed using standard benchmarks, such as the tasks from the Language Model Evaluation Harness~\cite{eval-harness}, including Arc-Easy, HellaSwag, PIQA and Winograde, which probe reasoning and generalization across a variety of domains. Additionally, we tested on the LAMBADA dataset~\cite{paperno2016lambada}, a word-prediction/ cloze-style natural language understanding task. We extended our evaluation to include Vision-Mamba models~\cite{hatamizadeh2025Mambavision}, such as Mambavision-S-1K and Mambavision-L-21K in half-precision (16-bit floating-point or FP16) format trained on ImageNet data~\citep{deng2009imagenet} to showcase multi-modal effectiveness of COBRA. Furthermore, we assess COBRA on Hymba~\cite{dong2024hymba}, a hybrid-head parallel architecture that combines transformer attention mechanisms and SSMs, to show the versatility of the proposed framework.  We report both \textbf{perplexity} (on WikiText~\cite{hu2024can}) and \textbf{accuracy} as evaluation metrics. Perplexity, defined as the exponential of the average negative log-likelihood over a sequence, measures predictive capability~\cite{hu2024can}, whereas accuracy quantifies the proportion of correct predictions.
\begin{table*}[t!]
\caption{COBRA evaluation on various models and datasets.}
\vspace{-2mm}
\label{table:main}
\resizebox{\textwidth}{!}{\begin{tabular}{c|c|ccccccc}\hline
\rowcolor[HTML]{EFEFEF} 
\cellcolor[HTML]{EFEFEF}                                 & \cellcolor[HTML]{EFEFEF}                                        & \multicolumn{7}{c}{\cellcolor[HTML]{EFEFEF}\textbf{Benchmarks (WikiText perplexity and \% Accuracy before/after attack)}}                                                                                                                                                                                                                                                                                                                                                                                                       \\ \cline{3-9} 
\rowcolor[HTML]{EFEFEF} 
\multirow{-2}{*}{\cellcolor[HTML]{EFEFEF}\textbf{Model}} & \multirow{-2}{*}{\cellcolor[HTML]{EFEFEF}\textbf{\# Bit-Flips}} & \multicolumn{1}{c|}{\cellcolor[HTML]{EFEFEF}\textbf{\begin{tabular}[c]{@{}c@{}}Perplexity\end{tabular}}} & \multicolumn{1}{c|}{\cellcolor[HTML]{EFEFEF}\textbf{Arc-Easy}} & \multicolumn{1}{c|}{\cellcolor[HTML]{EFEFEF}\textbf{Lambada}} & \multicolumn{1}{c|}{\cellcolor[HTML]{EFEFEF}\textbf{HellaSwag}} & \multicolumn{1}{c|}{\cellcolor[HTML]{EFEFEF}\textbf{PIQA}} & \multicolumn{1}{c|}{\cellcolor[HTML]{EFEFEF}\textbf{Winogrande}} & \textbf{ImageNet-1K} \\ \hline \hline
Mamba-370m                                               & \textbf{4}                                                               & \multicolumn{1}{c|}{24.87 / 1.26 x 10$^{11}$}                                                       & \multicolumn{1}{c|}{53.125\% / 28.67\%}                        & \multicolumn{1}{c|}{67.86\% / 5.33\%}                         & \multicolumn{1}{c|}{43.04\% / 22\%}                             & \multicolumn{1}{c|}{67.03\% / 44.66\%}                     & \multicolumn{1}{c|}{52.72\% / 46\%}                              & NA                   \\
Mamba2-370m                                              & \textbf{4}                                                            & \multicolumn{1}{c|}{26.69 / 2.48 x 10$^4$}                                                        & \multicolumn{1}{c|}{46.975\% / 12.67\%}                        & \multicolumn{1}{c|}{67.98\% / 17.33\%}                        & \multicolumn{1}{c|}{43.08\% / 8\%}                              & \multicolumn{1}{c|}{68.77\% / 24\%}                        & \multicolumn{1}{c|}{51.46\% / 7.33\%}                            & NA                    \\ \hline
Mamba-1.4b                                               & \textbf{1 }                                                              & \multicolumn{1}{c|}{18.94 / 3.75 x 10$^6$}                                                        & \multicolumn{1}{c|}{62.5\% / 14\%}                             & \multicolumn{1}{c|}{74.64\% / 0\%}                            & \multicolumn{1}{c|}{55.76\% / 0\%}                              & \multicolumn{1}{c|}{73.34\% / 18\%}                        & \multicolumn{1}{c|}{54.54\% / 46\%}                              & NA                  \\
Quamba-1.4b-w8a8                                         & \textbf{4}                                                               & \multicolumn{1}{c|}{22.193 / 50.08}                                                                                  & \multicolumn{1}{c|}{63.15\% / 24\%}                            & \multicolumn{1}{c|}{54.67\% / 46.67\%}                        & \multicolumn{1}{c|}{49.39\% / 22.66\%}                          & \multicolumn{1}{c|}{71.76\% / 49.33\%}                     & \multicolumn{1}{c|}{53.51\% / 27.33\%}                           & NA                    \\
Quamba-1.4b-w4a16                                        & \textbf{1}                                                               & \multicolumn{1}{c|}{23.62 / 606.673}                                                                                 & \multicolumn{1}{c|}{61.75\% / 0\%}                             & \multicolumn{1}{c|}{73.39\% / 57.33\%}                        & \multicolumn{1}{c|}{55.17\% / 0\%}                              & \multicolumn{1}{c|}{72.47\% / 6\%}                         & \multicolumn{1}{c|}{53.57\% / 6.67\%}                            & NA                    \\
Mamba2-1.3b                                              & \textbf{5}                                                              & \multicolumn{1}{c|}{18.76 / 2.41 x 10$^6$}                                                        & \multicolumn{1}{c|}{62.5\% / 0\%}                              & \multicolumn{1}{c|}{75.39\% / 30.67\%}                        & \multicolumn{1}{c|}{56.69\% / 0\%}                              & \multicolumn{1}{c|}{72.25\% / 0\%}                         & \multicolumn{1}{c|}{49.33\% / 4.67\%}                            & NA                   \\ \hline
Mamba-2.8b                                               & \textbf{1}                                                               & \multicolumn{1}{c|}{16.31 / 1.23 x 10$^5$}                                                        & \multicolumn{1}{c|}{65.625\% / 26.67\%}                        & \multicolumn{1}{c|}{78.34\% / 4\%}                            & \multicolumn{1}{c|}{62.59\% / 26\%}                             & \multicolumn{1}{c|}{74.48\% / 52\%}                        & \multicolumn{1}{c|}{56.99\% / 44.67\%}                           & NA                    \\
Quamba-2.8b-w8a8                                         & \textbf{1}                                                               & \multicolumn{1}{c|}{19.53 / 1.26x10$^{10}$}                                                         & \multicolumn{1}{c|}{65.26\% / 0\%}                             & \multicolumn{1}{c|}{77.55\% / 22\%}                           & \multicolumn{1}{c|}{62.06\% / 0\%}                              & \multicolumn{1}{c|}{73.99\% / 0\%}                         & \multicolumn{1}{c|}{56.59\% / 0\%}                               & NA                    \\
Quamba-2.8b-w4a16                                        & \textbf{1}                                                               & \multicolumn{1}{c|}{9.91 / 7.65 x 10$^5$}                                                         & \multicolumn{1}{c|}{68.7\% / 60\%}                             & \multicolumn{1}{c|}{76.03\% / 0\%}                            & \multicolumn{1}{c|}{61.66\% / 0\%}                              & \multicolumn{1}{c|}{73.29\% / 25.33\%}                     & \multicolumn{1}{c|}{58.01\% / 46\%}                              & NA                    \\
Mamba2-2.7b                                              & \textbf{9}                                                              & \multicolumn{1}{c|}{16.59 / 2.48 x 10$^4$}                                                        & \multicolumn{1}{c|}{66.41\% / 2\%}                             & \multicolumn{1}{c|}{77.78\% / 15.33\%}                        & \multicolumn{1}{c|}{62.69\% / 1.33\%}                           & \multicolumn{1}{c|}{74.86\% / 17.33\%}                     & \multicolumn{1}{c|}{56.59\% / 5.33\%}                            & NA                    \\ \hline
Hymba-1.5b                                               & \textbf{2}                                                               & \multicolumn{1}{c|}{14.40 / 1.762 x 10$^4$}                                                       & \multicolumn{1}{c|}{76.94\% / 0\%}                             & \multicolumn{1}{c|}{82.20\% / 30\%}                           & \multicolumn{1}{c|}{53.55\% / 0.67\%}                           & \multicolumn{1}{c|}{77.31\% / 2.67\%}                      & \multicolumn{1}{c|}{66.61\% / 10.67\%}                           & NA              \\
\hline
Mambavision-S-1K 50m                                          &     \textbf{22}                                                            & \multicolumn{1}{c|}{NA}                                                                                               & \multicolumn{1}{c|}{NA}                                         & \multicolumn{1}{c|}{NA}                                        & \multicolumn{1}{c|}{NA}                                          & \multicolumn{1}{c|}{NA}                                     & \multicolumn{1}{c|}{NA}                                           &   83.2\%/46.1\%    \\
Mambavision-L-21K 200m                                           &      \textbf{24}                                                           & \multicolumn{1}{c|}{NA}                                                                                               & \multicolumn{1}{c|}{NA}                                         & \multicolumn{1}{c|}{NA}                                        & \multicolumn{1}{c|}{NA}                                          & \multicolumn{1}{c|}{NA}                                     & \multicolumn{1}{c|}{NA}                                           &   86.1\%/47.2\%    \\
\hline
\end{tabular}
}
\end{table*}
\subsection{Preliminary Analysis}
\subsubsection{Layer Sensitivity Analysis}
\begin{figure}[t!]
  \centering
  \begin{subfigure}{0.48\linewidth}
    \includegraphics[width=1\linewidth]{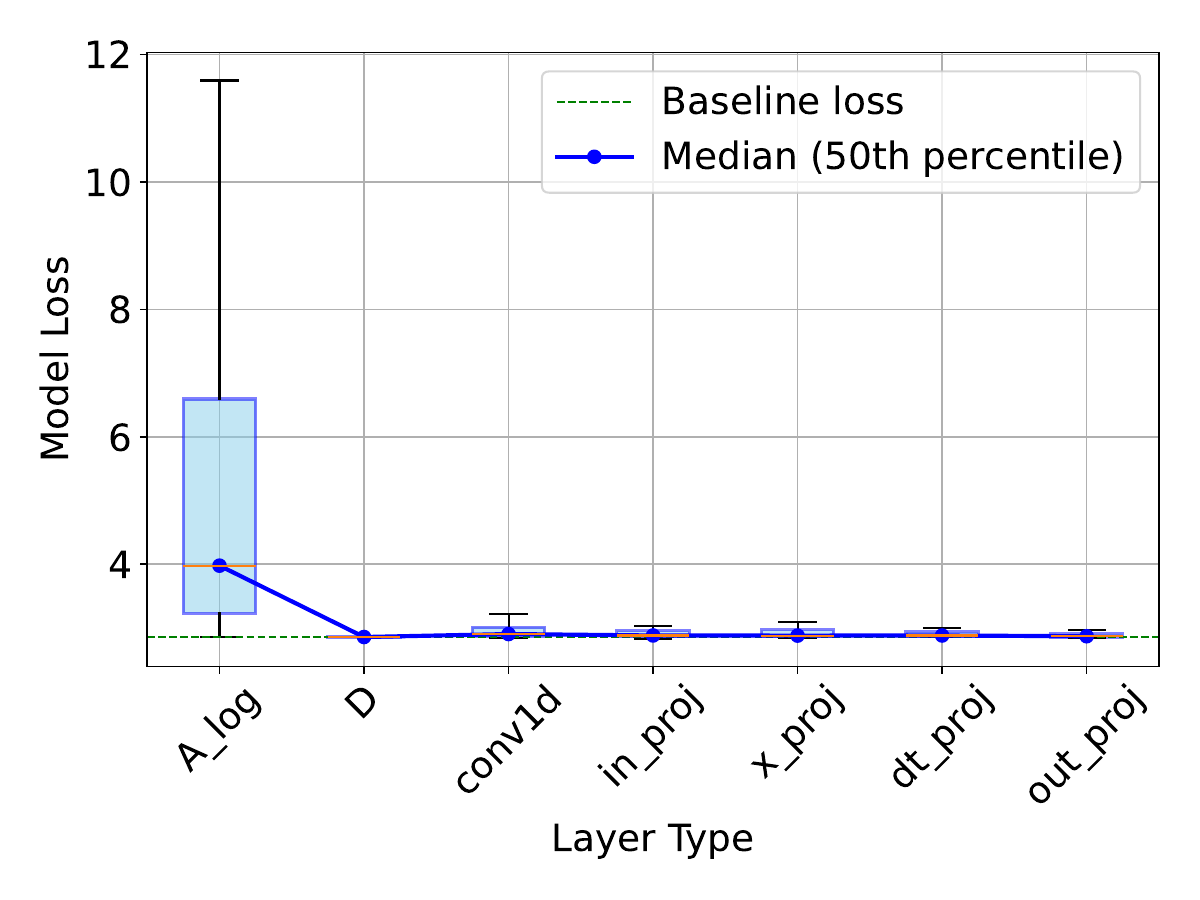}
    \caption{}
    \label{fig:ltype1}
  \end{subfigure}
\begin{subfigure}{0.48\linewidth}
    \includegraphics[width=1\linewidth]
{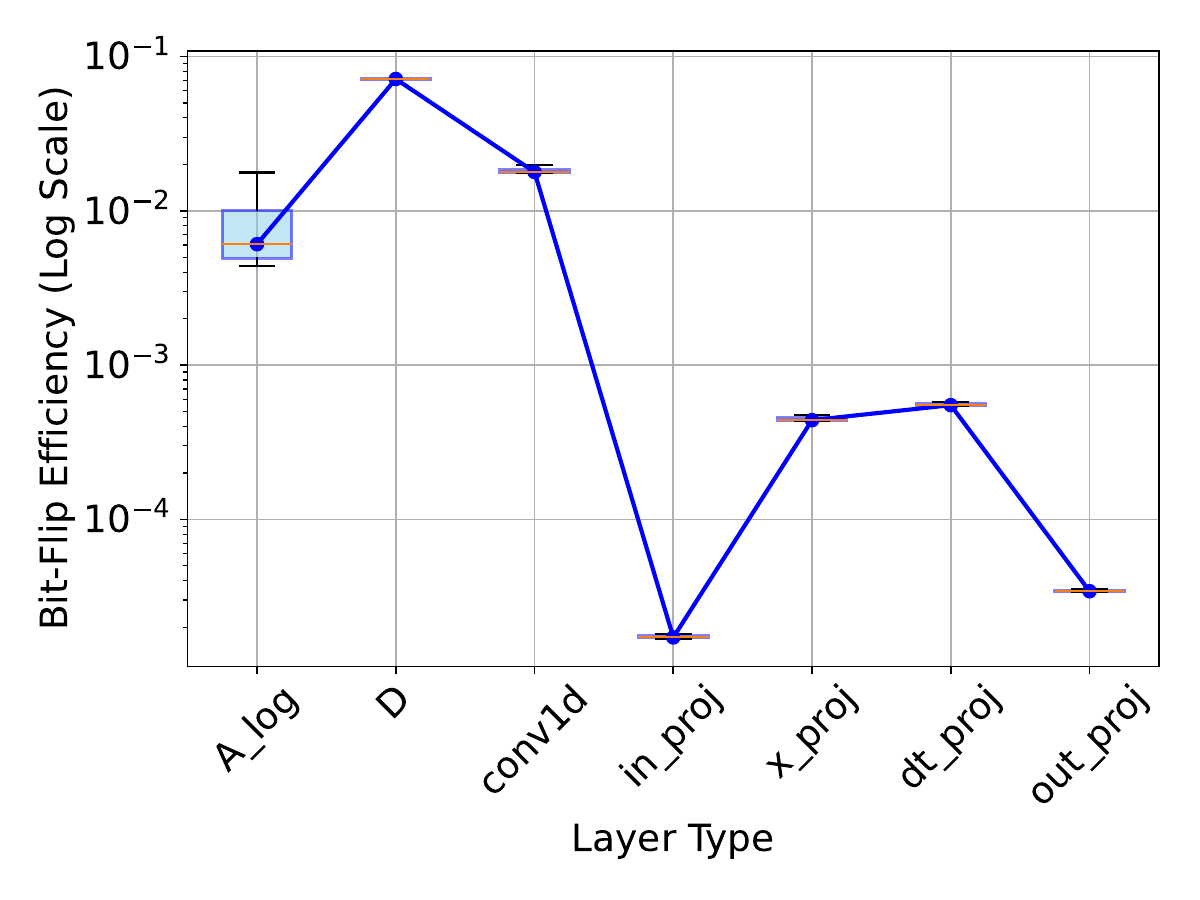}
    \caption{}
    \label{fig:ltype2}
  \end{subfigure}
  \caption{Layer type-based sensitivity analysis (a) loss distribution, (b) bit-flip efficiency in Mamba 1.4b FP16 model.}
  \label{fig:int_findings}
  \vspace{-2mm}
\end{figure}

\blue{We perform bit-flip injections at a fixed rate to quantify how different layer types affect model degradation (loss increase and consequent accuracy drop). Figure \ref{fig:int_findings} summarizes these results. Figure \ref{fig:ltype1} reports the absolute increase in model loss caused by bit-flips per layer type in Mamba-1.4b model upon the introduction of $0.1$\% bit-flips in ranked critical parameters (refer Section \ref{sec:proxy}). Larger losses indicate greater criticality. It is observed that the layer type $A_{log} $ is more critical compared to other layer types as bit-flips in these layers result in higher model losses and large accuracy degradation. This can be ascribed to the fact that ($A_{\log}$) directly parameterizes the state-transition matrix (A) through an exponential mapping, so perturbations in ($A_{\log}$) are amplified in (A), leading to large model loss, as discussed in Section~\ref{sec:theory}.} 



\blue{Layer sensitivity must also account for bit-flip efficiency, as layer sizes vary significantly and raw loss alone is not a sufficient indicator. We define bit-flip efficiency as the loss increase per flipped bit (Figure \ref{fig:ltype2}); higher values denote greater impact per perturbation. Layers exhibiting both high raw loss upon bit-flips and high efficiency are the most favorable BFA targets. Consistent with the theoretical analysis in Section \ref{sec:theory}, the $A_{log}$ layers in the Mamba/Quamba, $D$ and input and output projection layers in Mamba2 and $Conv1d$ layers in Hymba and Mambavision models show the highest criticality and efficiency, making them the most sensitive layer types.
}

\subsubsection{Weight-bit Subset Selection}
\begin{figure}[t!]
\vspace{-3mm}
  \centering
  \begin{subfigure}{0.48\linewidth}
    \includegraphics[width=1.1\linewidth]{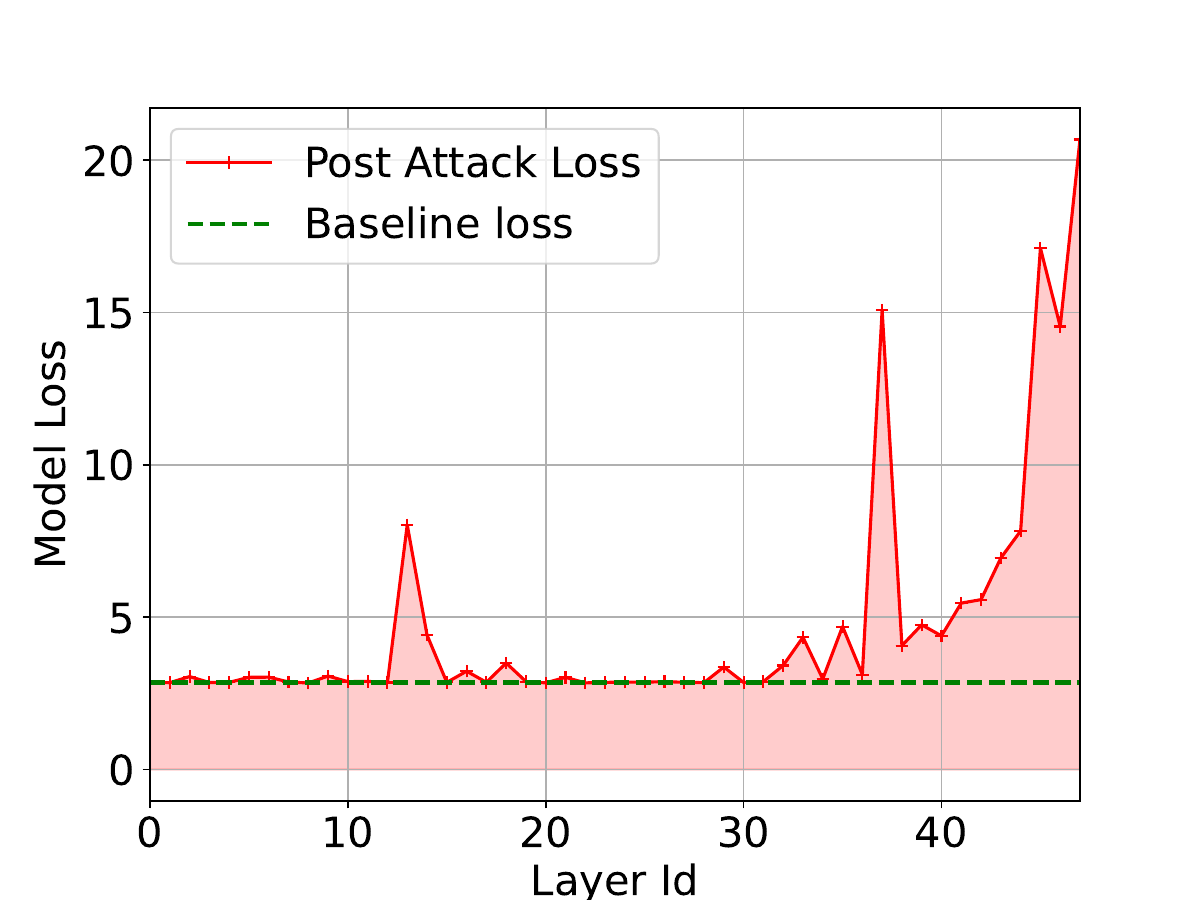}
    \caption{}
    \label{fig:bstype1}
  \end{subfigure}
\begin{subfigure}{0.48\linewidth}
    \includegraphics[width=1.1\linewidth]
{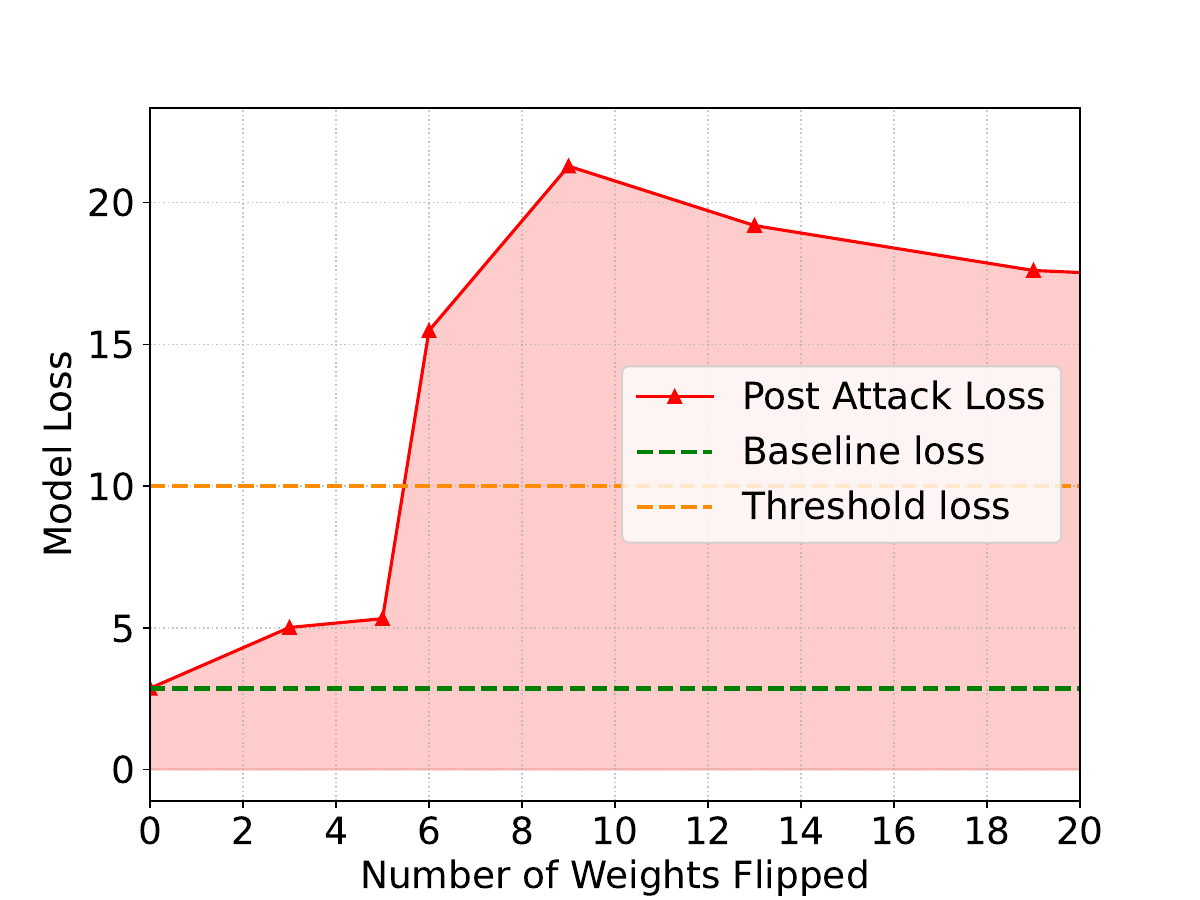}
    \caption{}
    \label{fig:bstype2}
  \end{subfigure}
  \caption{Critical (a) layer, and (b) weight subset selection in Mamba 1.4b FP16 model.}
  \label{fig:wsselection}
  \vspace{-3mm}
\end{figure}

\blue{We select the most-sensitive layer type identified in the preceding analysis (\textit{e.g.}, the $A_{log}$ layer in Mamba-1.4b) and determine a critical subset of parameters whose bit-flip perturbations yield a substantially high model loss.} Model loss (y-axis), as depicted in Figure \ref{fig:bstype1}, increases progressively with fixed $0.1$\% bit-flips in A\_log layers, in Mamba-1.4b model, from the initial to final Mamba blocks (x-axis). This indicates an increase in layer criticality from initial to final blocks.
Therefore, we target the A\_log layer in the final Mamba block for attack. Figure \ref{fig:bstype2} shows loss (y-axis) as a function of injected bit-flips (x-axis): the loss surpasses our predefined threshold of 10 after six bit-flips. We conservatively use the parameter set identified at a operating point of 9 bit-flips, (loss $\ge$ 21) as the initial critical weight subset for the subsequent exclusionary optimization.

\subsubsection{Weight-bit Subset Optimization}
\begin{figure}[b!]
  \centering
  \vspace{-4mm}
  \begin{subfigure}{0.49\linewidth}
    \includegraphics[width=1\linewidth]{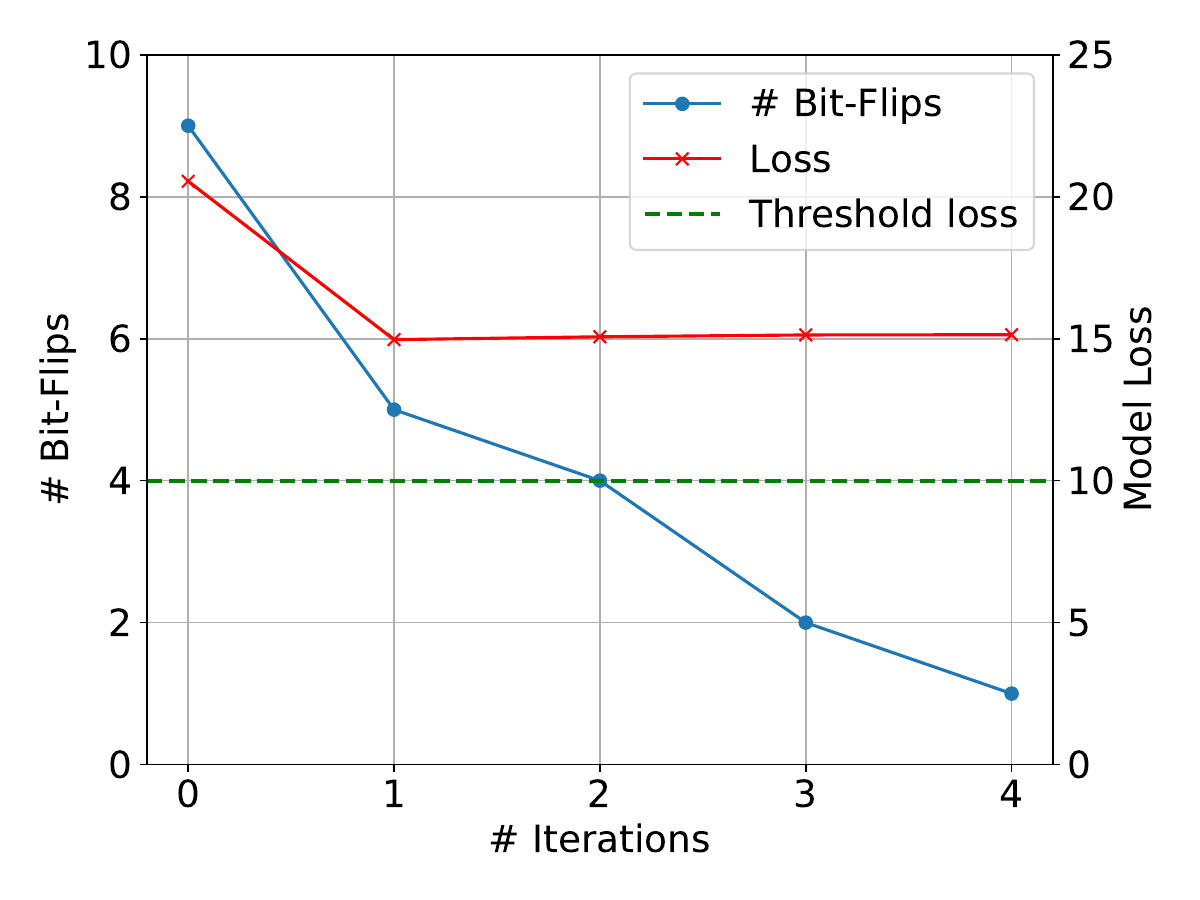}
    \caption{}
    \label{fig:bsopt1}
  \end{subfigure}
\begin{subfigure}{0.49\linewidth}
    \includegraphics[width=1\linewidth]
{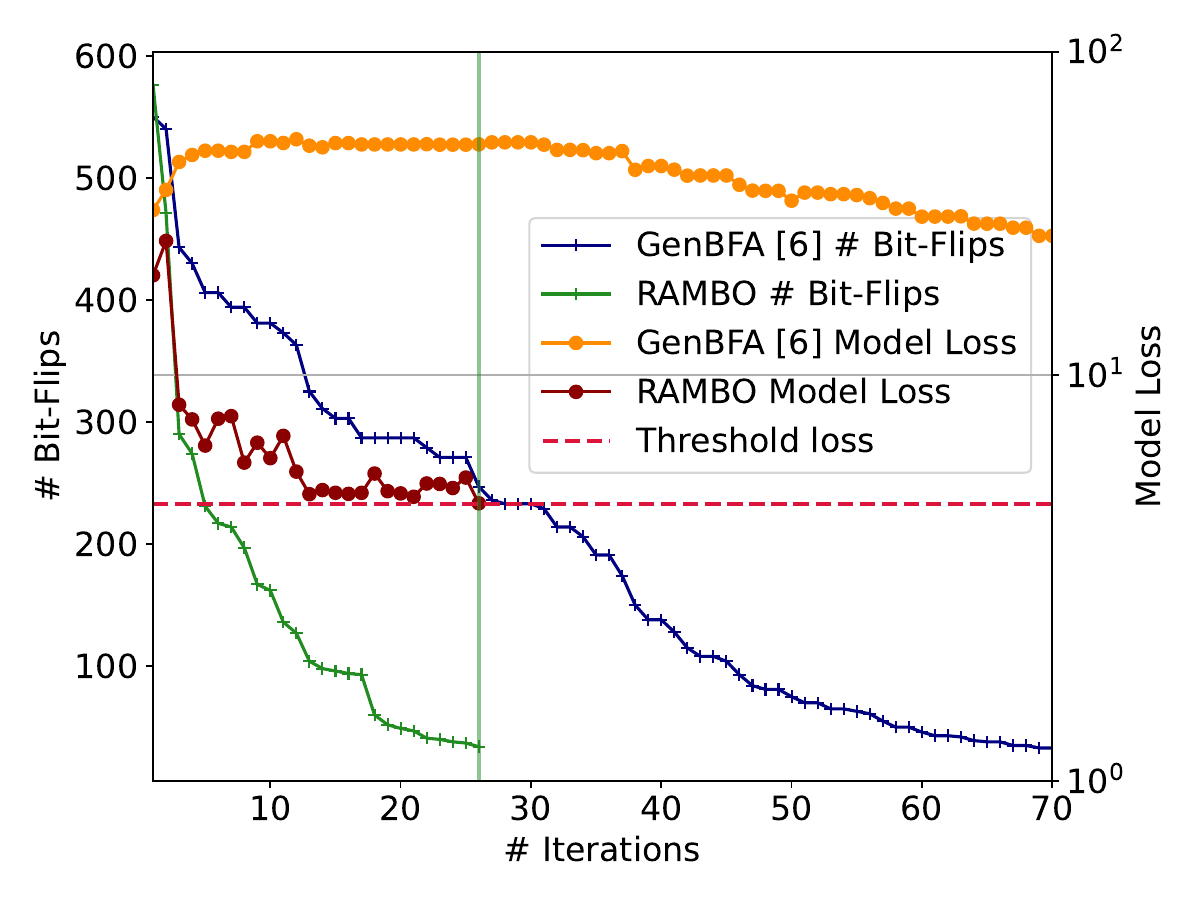}
    \caption{}
    \label{fig:bsopt2}
  \end{subfigure}
  \caption{(a) Weight-bit set optimization in Mamba 1.4b FP16 model, and (b) optimization perfromance comparison with GenBFA of AttentionBreaker framework\cite{das2024attentionbreaker} on Mambavision-S-1K.}
  \label{fig:wsopt}
\end{figure}

In this section, we refine the previously identified critical weight-bit subset to isolate the most critical bits. As shown in Figure \ref{fig:bsopt1}, the bit-flip attack optimization in the Mamba-1.4b FP16 model reduces the initial 9-bit subset to a \textbf{single} bit, indicated by the blue line, while maintaining a model loss (red line) above the loss threshold of 10 (green dotted line). This demonstrates the necessity and effectiveness of the optimization process in identifying most critical bits.

\subsection{Results}

This segment presents the degradation of model performance across benchmarks achieved by COBRA (Table \ref{table:main}). The attributes of each model, specifically the model name and parameter count, are presented in the firts column. The second column presents the count of bit-flips injected to induce performance degradation. Subsequent columns furnish benchmark results, before and after the attack, for each task. 
All models demonstrated strong baseline performance, with LAMBADA accuracies up to 82.20\%, and WikiText perplexities between 16.31 and 26.69. Similarly, high accuracies were observed across reasoning and commonsense benchmarks such as ARC-Easy (53.12– 76.94\%), HellaSwag (43.04–62.69\%), PIQA (67.03–77.31\%), and Winogrande (51.46– 66.61\%). ``NA'' indicates when the benchmark is incompatible and does not
apply to the model.

Following the injection of only a few targeted \textbf{bit-flips}, COBRA produced severe accuracy degradation across all tasks. In the \textbf{Mamba-1.4b} model, a \textbf{single bit-flip} caused complete collapse on LAMBADA (accuracy dropped from 74.64\% to 0\%) and raised perplexity from 18.94 to 3.75 × 10$^6$, while also reducing ARC-Easy accuracy from 62.5\% to 14\%. The \textbf{Mamba-2.8b }model exhibited similar vulnerability, where \textbf{one} bit-flip reduced LAMBADA accuracy from 78.34\% to 4\% and increased perplexity from 16.31 to 1.23 × 10$^5$. Even smaller models, such as Mamba-370M, experienced notable drops, with four bit-flips lowering Arc-Easy accuracy from 53.125\% to 28.67\% and increasing perplexity from 24.87 to 1.26 $\times 10^{11}$. 

Similarly, Mamba2 are also highly sensitive: Mamba2-1.3b required 5 bit-flips to collapse Arc-Easy accuracy to 0\%, while Mamba2-2.7B requires 32 bit-flips to induce similar attack impact.
\textit{Overall, these results reveal that Mamba architectures exhibit extreme bit-level fragility, where even \textbf{a single bit-flip} in critical parameters can induce catastrophic failures in model performance.}
\subsection{Attack Efficacy across Quantization Levels}
Our evaluation has, up to this point, concentrated on COBRA using half-precision FP16 models. However, a model’s vulnerability to bit-flip attacks is largely determined by quantization precision. Since diverse numerical formats display varying sensitivities to perturbations, the effect of a bit-flip, along with the quantity of flips needed to cause model disruption, can differ significantly across different quantizations~\cite{das2024attentionbreaker}. Furthermore, previous BFA defenses commonly suggest quantization as a strategy for mitigation, necessitating the evaluation of COBRA under quantized conditions~\cite{rakin2021t}.

To this end, we evaluate COBRA on INT4 (w4a16) and INT8 (w8a8) variants of Mamba-1.4b and Mamba-2.8b. Our findings show that these quantized models remain highly vulnerable to bit-flip attacks. 
Notably, a \textbf{single} bit-flip is sufficient to reduce the accuracy of the Mamba-2.8b w4a16 model on PIQA from 73.29\% to 25.33\% (refer Table \ref{table:main}), demonstrating that COBRA remains extremely effective even under aggressive quantization.
\subsection{Attack Efficacy across Task Modality}
To assess COBRA’s effectiveness beyond NLP settings, we applied it to FP16 Mamba-Vision models with 50M–200M parameters trained on ImageNet. Our results show that COBRA remains highly effective even in vision tasks, successfully identifying critical bit positions whose perturbation leads to substantial performance degradation. For example, in the MambaVision-L-21K model, COBRA identified 24 critical bits that, when flipped, reduced the model’s accuracy from 86.1\% to 47.2\% (refer Table \ref{table:main}). This demonstrates that COBRA generalizes effectively beyond NLP and remains a potent bit-flip attack methodology for Mamba-based architectures on vision tasks.

\subsection{Attack Transferability across Benchmarks}
This experiment evaluates the cross-task transferability of bit‐flip attack effects. Specifically, we examine whether an attack designed to degrade performance on Task A also induces comparable degradation on Task B. Such transferability would indicate a fundamental architectural vulnerability that is independent of the specific task.

Using the Mamba-1.4b FP16 model, we first execute a single bit-flip attack on the ARC-Easy benchmark, reducing accuracy from 62.5\% to 14\%. We then measure the resulting impact on other language benchmarks, including HellaSwag, PIQA, and Winogrande. As summarized in Table \ref{table:main}, the attack exhibits substantial transferability: for example, HellaSwag accuracy drops from 55.76\% to 0\%, and PIQA accuracy declines from 73.34\% to 18\%. These results demonstrate that the effects of the attack propagate broadly across tasks and domains, underscoring a fundamental and widespread vulnerability in the model architecture.

\subsection{Gradient-free Attack Results}
We evaluate COBRA in a fully gradient-free setting, where gradient information is unavailable throughout the attack. Parameter sensisitivity scores are derived solely from weight magnitudes by setting $\alpha = 0$ in Equation \ref{eq:sensitivity}. Under this configuration, only a \textbf{single} bit-flip on the 8-bit Quamba-2.8B-W8A8 model reduces ARC-Easy accuracy from 65.26\% to 0\%, demonstrating that magnitude-based scoring remains highly effective in identifying critical bits. This flexibility allows adversaries to select either gradient-free or gradient-based modes depending on their goals and computational constraints.

Notably, many existing BFA defenses rely on restricting or obfuscating gradient access. COBRA circumvents such defenses entirely by operating on magnitude-based importance alone. This adaptability underscores COBRA’s robustness in gradient-restricted environments and establishes it as a resilient BFA framework.

\subsection{Gray-box Attack Results}
To evaluate COBRA's effectiveness under grey-box scenarios with partial access to model parameters, we simulate such a condition by restricting access to a only a subset of model layers, specifically t\textbf{he final two Mamba blocks} in the Mamba-1.4b model. 
Even with this limited access, COBRA remains highly potent. On the Arc-Easy dataset, \textbf{flipping only a single bit} in this partially visible Mamba-1.4b model is sufficient to reduce accuracy from 62.5\% to 15.2\%, demonstrating the strength of our approach under constrained access.

\subsection{Comparison with State-of-the-art LLM Bit-flip Attacks}
Although COBRA is the first SSM-aware BFA and thus lacks a direct SSM-specific baseline, we assess how its exclusionary optimization strategy compares with LLM-based BFAs. To this end, we evaluate it against GenBFA optimization employed in the  AttentionBreaker framework~\cite{das2024attentionbreaker}. As shown in Figure \ref{fig:bsopt2}, we observe that both methods identify the same 22 critical bits in Mambavision-S-1K, but COBRA converges significantly faster, within 26 iterations (green line) compared to GenBFA’s 70 (blue line), demonstrating superior efficiency despite its simpler design.


\blue{Prior works, such as AttentionBreaker~\cite{das2024attentionbreaker} and SBFA~\cite{guo2025sbfa}, show that \textbf{single or few bit-flips} can collapse models with billions of parameters. Similarly, COBRA achieves catastrophic degradation with just \textbf{one} bit-flip in Mamba-1.4b (FP16) and Quamba-2.8b (INT4). Even the hybrid Hymba architecture is oberserved to be extremely vulnerable and its performance colapses with merely \textbf{two} bit-flips. 
These results highlight the severe security risk posed by bit-flip attacks in both transformer- and state-space-based models.}

\section{Conclusion}\label{sec:conclusion}
This research investigates the vulnerability of state-space based models like Mamba and presents COBRA, a novel BFA framework targeting these models. COBRA introduces a novel sensitivity analysis of state space model architectures, thereby resulting in compromised performance. Specifically, perturbing merely one-bit (7.14 x 10$^{-10}$\% of the model parameters) in Mamba-1.4b model results on LAMBADA dataset prediction scores dropping from 74.64\% to 0\%. The results highlight COBRA’s utility in demonstrating Mamba’s vulnerability to such adversarial interventions.
\section{Ethical Considerations}
This study aims to advance research on the safety and robustness of NLP models, with a particular focus on state-space architectures such as Mamba. It is imperative that the hardware fault vulnerabilities and attack methodologies presented here are used solely for constructive purposes, and not for malicious exploitation.
\balance

\bibliography{main}

@inproceedings{mao2023cross,
  title={Cross-entropy loss functions: Theoretical analysis and applications},
  author={Mao, Anqi and Mohri, Mehryar and Zhong, Yutao},
  booktitle={International conference on Machine learning},
  pages={23803--23828},
  year={2023},
  organization={pmlr}
}

@article{kim2014flipping,
  title={Flipping bits in memory without accessing them: An experimental study of DRAM disturbance errors},
  author={Kim, Yoongu and Daly, Ross and Kim, Jeremie and Fallin, Chris and Lee, Ji Hye and Lee, Donghyuk and Wilkerson, Chris and Lai, Konrad and Mutlu, Onur},
  journal={ACM SIGARCH Computer Architecture News},
  volume={42},
  number={3},
  pages={361--372},
  year={2014},
  publisher={ACM New York, NY, USA}
}

@inproceedings{kayaalp2016high,
  title={A high-resolution side-channel attack on last-level cache},
  author={Kayaalp, Mehmet and Abu-Ghazaleh, Nael and Ponomarev, Dmitry and Jaleel, Aamer},
  booktitle={Proceedings of the 53rd Annual Design Automation Conference},
  pages={1--6},
  year={2016}
}

@inproceedings{saha2020hidden,
  title={Hidden trigger backdoor attacks},
  author={Saha, Aniruddha and Subramanya, Akshayvarun and Pirsiavash, Hamed},
  booktitle={Proceedings of the AAAI conference on artificial intelligence},
  volume={34},
  pages={11957--11965},
  year={2020}
}

@inproceedings{xie2019dba,
  title={Dba: Distributed backdoor attacks against federated learning},
  author={Xie, Chulin and Huang, Keli and Chen, Pin-Yu and Li, Bo},
  booktitle={International Conference on Learning Representations},
  year={2019}
}

@article{goodfellow2014explaining,
  title={Explaining and harnessing adversarial examples},
  author={Goodfellow, Ian J and Shlens, Jonathon and Szegedy, Christian},
  journal={arXiv preprint arXiv:1412.6572},
  year={2014}
}

@inproceedings{moosavi2017universal,
  title={Universal adversarial perturbations},
  author={Moosavi-Dezfooli, Seyed-Mohsen and Fawzi, Alhussein and Fawzi, Omar and Frossard, Pascal},
  booktitle={Proceedings of the IEEE conference on computer vision and pattern recognition},
  pages={1765--1773},
  year={2017}
}

@inproceedings{breier2018practical,
  title={Practical fault attack on deep neural networks},
  author={Breier, Jakub and Hou, Xiaolu and Jap, Dirmanto and Ma, Lei and Bhasin, Shivam and Liu, Yang},
  booktitle={Proceedings of the 2018 ACM SIGSAC Conference on Computer and Communications Security},
  pages={2204--2206},
  year={2018}
}

@inproceedings{selmke2015precise,
  title={Precise laser fault injections into 90 nm and 45 nm sram-cells},
  author={Selmke, Bodo and Brummer, Stefan and Heyszl, Johann and Sigl, Georg},
  booktitle={International Conference on Smart Card Research and Advanced Applications},
  pages={193--205},
  year={2015},
  organization={Springer}
}

@inproceedings{deng2009imagenet,
  title={Imagenet: A large-scale hierarchical image database},
  author={Deng, Jia and Dong, Wei and Socher, Richard and Li, Li-Jia and Li, Kai and Fei-Fei, Li},
  booktitle={2009 IEEE conference on computer vision and pattern recognition},
  pages={248--255},
  year={2009},
  organization={Ieee}
}

@article{mamba,
  title={Mamba: Linear-Time Sequence Modeling with Selective State Spaces},
  author={Gu, Albert and Dao, Tri},
  journal={arXiv preprint arXiv:2312.00752},
  year={2023}
}

@inproceedings{paperno2016lambada,
  title={The LAMBADA dataset: Word prediction requiring a broad discourse context},
  author={Paperno, Denis and Kruszewski, Germ{\'a}n and Lazaridou, Angeliki and Pham, Ngoc-Quan and Bernardi, Raffaella and Pezzelle, Sandro and Baroni, Marco and Boleda, Gemma and Fern{\'a}ndez, Raquel},
  booktitle={Proceedings of the 54th annual meeting of the association for computational linguistics (volume 1: Long papers)},
  pages={1525--1534},
  year={2016}
}

@inproceedings{mamba2,
  title={Transformers are {SSM}s: Generalized Models and Efficient Algorithms Through Structured State Space Duality},
  author={Dao, Tri and Gu, Albert},
  booktitle={International Conference on Machine Learning (ICML)},
  year={2024}
}

@article{gu2023mamba,
  title={Mamba: Linear-time sequence modeling with selective state spaces},
  author={Gu, Albert and Dao, Tri},
  journal={arXiv preprint arXiv:2312.00752},
  year={2023}
}

@article{das2024attentionbreaker,
  title={Attentionbreaker: Adaptive evolutionary optimization for unmasking vulnerabilities in llms through bit-flip attacks},
  author={Das, Sanjay and Bhattacharya, Swastik and Kundu, Souvik and Kundu, Shamik and Menon, Anand and Raha, Arnab and Basu, Kanad},
  journal={arXiv e-prints},
  pages={arXiv--2411},
  year={2024}
}

@article{qian2023survey,
  title={A survey of bit-flip attacks on deep neural network and corresponding defense methods},
  author={Qian, Cheng and Zhang, Ming and Nie, Yuanping and Lu, Shuaibing and Cao, Huayang},
  journal={Electronics},
  volume={12},
  number={4},
  pages={853},
  year={2023},
  publisher={MDPI}
}

@inproceedings{codematch,
  title={Improving robustness against stealthy weight bit-flip attacks by output code matching},
  author={{\"O}zdenizci, Ozan and Legenstein, Robert},
  booktitle={Proceedings of the IEEE/CVF Conference on Computer Vision and Pattern Recognition},
  pages={13388--13397},
  year={2022}
}

@inproceedings{yarom2014flush+,
  title={$\{$FLUSH+ RELOAD$\}$: A high resolution, low noise, l3 cache $\{$Side-Channel$\}$ attack},
  author={Yarom, Yuval and Falkner, Katrina},
  booktitle={23rd USENIX security symposium (USENIX security 14)},
  pages={719--732},
  year={2014}
}

@article{mahmood2021back,
  title={Back in black: A comparative evaluation of recent state-of-the-art black-box attacks},
  author={Mahmood, Kaleel and Mahmood, Rigel and Rathbun, Ethan and van Dijk, Marten},
  journal={IEEE Access},
  volume={10},
  pages={998--1019},
  year={2021},
  publisher={IEEE}
}

@inproceedings{papernot2017practical,
  title={Practical black-box attacks against machine learning},
  author={Papernot, Nicolas and McDaniel, Patrick and Goodfellow, Ian and Jha, Somesh and Celik, Z Berkay and Swami, Ananthram},
  booktitle={Proceedings of the 2017 ACM on Asia conference on computer and communications security},
  pages={506--519},
  year={2017}
}

@article{xiang2020side,
  title={Side-channel gray-box attack for dnns},
  author={Xiang, Yun and Xu, Yongchao and Li, Yingjie and Ma, Wen and Xuan, Qi and Liu, Yi},
  journal={IEEE Transactions on Circuits and Systems II: Express Briefs},
  volume={68},
  number={1},
  pages={501--505},
  year={2020},
  publisher={IEEE}
}

@article{madry2017towards,
  title={Towards deep learning models resistant to adversarial attacks},
  author={Madry, Aleksander},
  journal={arXiv preprint arXiv:1706.06083},
  year={2017}
}

@inproceedings{hatamizadeh2025mambavision,
  title={Mambavision: A hybrid mamba-transformer vision backbone},
  author={Hatamizadeh, Ali and Kautz, Jan},
  booktitle={Proceedings of the Computer Vision and Pattern Recognition Conference},
  pages={25261--25270},
  year={2025}
}

@article{dong2024hymba,
  title={Hymba: A hybrid-head architecture for small language models},
  author={Dong, Xin and Fu, Yonggan and Diao, Shizhe and Byeon, Wonmin and Chen, Zijia and Mahabaleshwarkar, Ameya Sunil and Liu, Shih-Yang and Van Keirsbilck, Matthijs and Chen, Min-Hung and Suhara, Yoshi and others},
  journal={arXiv preprint arXiv:2411.13676},
  year={2024}
}

@article{guo2025sbfa,
  title={SBFA: Single Sneaky Bit Flip Attack to Break Large Language Models},
  author={Guo, Jingkai and Chakrabarti, Chaitali and Fan, Deliang},
  journal={arXiv preprint arXiv:2509.21843},
  year={2025}
}

@article{rakin2021t,
  title={T-bfa: Targeted bit-flip adversarial weight attack},
  author={Rakin, Adnan Siraj and He, Zhezhi and Li, Jingtao and Yao, Fan and Chakrabarti, Chaitali and Fan, Deliang},
  journal={IEEE Transactions on Pattern Analysis and Machine Intelligence},
  volume={44},
  number={11},
  pages={7928--7939},
  year={2021},
  publisher={IEEE}
}

@inproceedings{rakin2019bit,
  title={Bit-flip attack: Crushing neural network with progressive bit search},
  author={Rakin, Adnan Siraj and He, Zhezhi and Fan, Deliang},
  booktitle={Proceedings of the IEEE/CVF International Conference on Computer Vision},
  pages={1211--1220},
  year={2019}
}

@article{xu2024survey,
  title={A survey of resource-efficient llm and multimodal foundation models},
  author={Xu, Mengwei and Yin, Wangsong and Cai, Dongqi and Yi, Rongjie and Xu, Daliang and Wang, Qipeng and Wu, Bingyang and Zhao, Yihao and Yang, Chen and Wang, Shihe and others},
  journal={arXiv preprint arXiv:2401.08092},
  year={2024}
}

@article{chang2024survey,
  title={A survey on evaluation of large language models},
  author={Chang, Yupeng and Wang, Xu and Wang, Jindong and Wu, Yuan and Yang, Linyi and Zhu, Kaijie and Chen, Hao and Yi, Xiaoyuan and Wang, Cunxiang and Wang, Yidong and others},
  journal={ACM Transactions on Intelligent Systems and Technology},
  volume={15},
  number={3},
  pages={1--45},
  year={2024},
  publisher={ACM New York, NY}
}

@inproceedings{chen2021proflip,
  title={Proflip: Targeted trojan attack with progressive bit flips},
  author={Chen, Huili and Fu, Cheng and Zhao, Jishen and Koushanfar, Farinaz},
  booktitle={Proceedings of the IEEE/CVF International Conference on Computer Vision},
  pages={7718--7727},
  year={2021}
}

@inproceedings{hayashi2011non,
  title={Non-invasive EMI-based fault injection attack against cryptographic modules},
  author={Hayashi, Yu-ichi and Homma, Naofumi and Sugawara, Takeshi and Mizuki, Takaaki and Aoki, Takafumi and Sone, Hideaki},
  booktitle={2011 IEEE International Symposium on Electromagnetic Compatibility},
  pages={763--767},
  year={2011},
  organization={IEEE}
}

@article{kundubit,
  title={Bit-by-Bit: Investigating the Vulnerabilities of Binary Neural Networks to Adversarial Bit Flipping},
  author={Kundu, Shamik and Das, Sanjay and Karmakar, Sayar and Raha, Arnab and Kundu, Souvik and Makris, Yiorgos and Basu, Kanad},
  journal={Transactions on Machine Learning Research},
  year={2024}
}

@inproceedings{yao2020deephammer,
  title={$\{$DeepHammer$\}$: Depleting the intelligence of deep neural networks through targeted chain of bit flips},
  author={Yao, Fan and Rakin, Adnan Siraj and Fan, Deliang},
  booktitle={29th USENIX Security Symposium (USENIX Security 20)},
  pages={1463--1480},
  year={2020}
}

@article{hu2024can,
  title={Can Perplexity Reflect Large Language Model's Ability in Long Text Understanding?},
  author={Hu, Yutong and Huang, Quzhe and Tao, Mingxu and Zhang, Chen and Feng, Yansong},
  journal={arXiv preprint arXiv:2405.06105},
  year={2024}
}

@misc{eval-harness,
  author       = {Gao, Leo and Tow, Jonathan and Abbasi, Baber and Biderman, Stella and Black, Sid and DiPofi, Anthony and Foster, Charles and Golding, Laurence and Hsu, Jeffrey and Le Noac'h, Alain and Li, Haonan and McDonell, Kyle and Muennighoff, Niklas and Ociepa, Chris and Phang, Jason and Reynolds, Laria and Schoelkopf, Hailey and Skowron, Aviya and Sutawika, Lintang and Tang, Eric and Thite, Anish and Wang, Ben and Wang, Kevin and Zou, Andy},
  title        = {A framework for few-shot language model evaluation},
  month        = 07,
  year         = 2024,
  publisher    = {Zenodo},
  version      = {v0.4.3},
  doi          = {10.5281/zenodo.12608602},
  url          = {https://zenodo.org/records/12608602}
}

@article{shuvo2023comprehensive,
  title={A comprehensive survey on non-invasive fault injection attacks},
  author={Shuvo, Amit Mazumder and Zhang, Tao and Farahmandi, Farimah and Tehranipoor, Mark},
  journal={Cryptology ePrint Archive},
  year={2023}
}

@article{das2024security,
  title={Security and privacy challenges of large language models: A survey},
  author={Das, Badhan Chandra and Amini, M Hadi and Wu, Yanzhao},
  journal={arXiv preprint arXiv:2402.00888},
  year={2024}
}

@inproceedings{nazari2024forget,
  title={Forget and Rewire: Enhancing the Resilience of Transformer-based Models against $\{$Bit-Flip$\}$ Attacks},
  author={Nazari, Najmeh and Makrani, Hosein Mohammadi and Fang, Chongzhou and Sayadi, Hossein and Rafatirad, Setareh and Khasawneh, Khaled N and Homayoun, Houman},
  booktitle={33rd USENIX Security Symposium (USENIX Security 24)},
  pages={1349--1366},
  year={2024}
}

@article{zhang2018generalized,
  title={Generalized cross entropy loss for training deep neural networks with noisy labels},
  author={Zhang, Zhilu and Sabuncu, Mert},
  journal={Advances in neural information processing systems},
  volume={31},
  year={2018}
}
\bibliographystyle{ACM-Reference-Format}
\end{document}